\documentclass[twocolumn,nofootinbib,showpacs,preprintnumbers,superscriptaddress] {revtex4}

\usepackage{amssymb,amsfonts,amsmath,graphicx,}
\usepackage{color}
\usepackage{enumerate}
\usepackage{braket}
\usepackage{epstopdf}
\usepackage{varioref}
\usepackage{slashed}

\newcommand{\bfp}{{\bf p}}

\newcommand{\dis}[1]{\begin{equation}\begin{split}#1\end{split}\end{equation}}

\newcommand{\bfrac}[2]{{\left(\frac{#1}{#2} \right)  }}
\newcommand{\eq}[1]{Eq.~(\ref{#1})}

\newcommand\tev{\,{\rm TeV}}
\newcommand\gev{\,{\rm GeV}}
\newcommand\mev{\,{\rm MeV}}

\newcommand\ev{\,{\rm eV}}

\newcommand\cm{\,{\rm cm}}

\newcommand\mnu{{m_\nu}}
\newcommand\mphi{m_\phi}
\newcommand\mf{m_f^2}
\newcommand\dmnu{\delta m_\nu^2}

\begin{document}

\title{Dispersion of neutrinos in a medium }

\author{Ki-Young Choi}
\email{kiyoungchoi@skku.edu}
 \affiliation{Department of Physics, Sungkyunkwan University,  16419 Korea}

\author{Eung Jin Chun}
\email{ejchun@kias.re.kr}
\affiliation{Korea Institute for Advanced Study, Seoul 02455, Korea}

\author{Jongkuk Kim}
\email{jkkim@kias.re.kr}
\affiliation{Korea Institute for Advanced Study, Seoul 02455, Korea}

\begin{abstract} 
We analyze the dispersion relations of Weyl or Majorana,  and Dirac neutrinos in a complex scalar medium which interacts with  the neutrinos through Yukawa couplings. They are solved by perturbative calculation in various limits representing different physical situations, some of which allow the medium-induced neutrino oscillation to occur. Remarkably, peculiar dispersion relations arise differently for Majorana or Dirac neutrinos in the non-relativistic limit. 
This provides an unpleasant restriction on the cosmological scenario of a scalar dark matter coupling to neutrinos. 
 At present, the model parameter space is constrained by the neutrino scattering with dark matter 
 through astrophysical neutrino observations.
\end{abstract}


\preprint{KIAS-P20072}

\maketitle

\section{ Introduction}
Coherent forward scattering of neutrinos propagating in background matter is shown to generate an effective (Wolfenstein) potential which modifies vacuum oscillations~\cite{Wolfenstein:1977ue}. In a medium of varying density, such an effect can lead to adiabatic neutrino conversion~\cite{Mikheev:1986gs} which is applied to solar neutrinos and confirmed by various experiments. 
This is in fact a particular feature of dispersion relation of fermions interacting with a medium of finite temperature and density~\cite{Weldon:1982bn}.
Such a technique of quantum field theory at finite temperature has been used to reproduce the Wolfenstein potential~\cite{ Mannheim:1987ef,Pal:1989xs,Notzold:1987ik,Nieves:1989ez,DOlivo:1992lwg,Kiers:1997yt}.

During the past decade, there was a great interest in considering various non-standard interactions of neutrinos in ordinary matter or dark matter (DM)
to study their consequences in neutrino oscillations as well as cosmological and astrophysics observations~\cite{Sawyer:1998ac,Farzan:2015hkd,Dev:2019anc,Venzor:2020ova,Berlin:2016woy,Brdar:2017kbt,Krnjaic:2017zlz,Liao:2018byh,Huang:2018cwo,Dev:2020kgz,Capozzi:2017auw,Capozzi:2018bps,Farzan:2018gtr,Smirnov:2019cae,Cline:2019seo}. For a systematic study of medium effect, one needs to analyze dispersion relations modified by non-standard interactions through  coherent forward scattering. Recently,  such an analysis was performed considering various thermal and DM backgrounds \cite{Nieves:2018vxl,Ge:2019tdi,Choi:2019zxy,Babu:2019iml}. In particular,  the first attempt to derive a DM potential modifying neutrino oscillations was made in~\cite{Ge:2019tdi,Choi:2019zxy}. It was noticed that neutrino oscillations can occur solely by the medium effect even for massless neutrinos~\cite{Choi:2019zxy}.

The purpose of this article is to provide a systematic analysis of dispersion relation of neutrinos propagating in a medium.  
Specifically, we consider Weyl or Majorana, and Dirac neutrinos interacting with a complex scalar medium  through a Yukawa coupling. 
The general equations of motion are highly non-linear and do not allow analytic solutions for the dispersion relation. 
However, they can be solved perturbatively considering various limits of physical interest with heavy mediator, heavy neutrino, high-momentum, and high density. Depending on the type of neutrinos, the effective mass and effective potential induced by the medium are different.  
From this analysis, one can see when the medium-induced oscillation occurs and how the asymmetric distribution of background field splits 
the energy spectrum of the neutrino and antineutrino. We also recognize the appearance of medium-induced mass-squared in the non-relativistic (or low-momentum) and high-density limit, which severely constrain the cosmological scenario of neutrino-DM interaction through CMB observations~\cite{Aghanim:2018eyx}. In the opposite limit of high-momentum, the astrophysical neutrino observations provide independent limits disfavoring a large part of parameter 
space~\cite{Choi:2019ixb}.

In Sec.~\ref{self-energy}, we provide the general formula for the self-energy correction of the neutrino in a medium. In Sec.~\ref{complex}, we find the approximate dispersion relation for each limiting case of Majorana and Dirac neutrinos. In Sec.~\ref{implications}, we consider the implications of the modified neutrino dispersions in a medium and conclude in Sec.~\ref{conclusions}.

\medskip

\section{Self-energy correction in a medium}
\label{self-energy}
For neutrinos propagating in a medium, the equations of motion are modified due to the forward elastic scattering with the background field, which is expressed in terms of the self-energy correction. In general, the corrections are different for  left-, and right-handed neutrinos as they may interact differently with the medium,  and also for neutrinos and antineutrinos when the background distribution is CP asymmetric.

 The left- and right-handed $u$ (or $v$) spinors in a medium satisfy the equations of motion
\dis{
\begin{matrix}
(\slashed{p}- \slashed{\Sigma}_L^u) u_L = M_\nu^\dagger u_R\\
(\slashed{p} - \slashed{\Sigma}_R^u) u_R = M_\nu u_L\\
\end{matrix}
~~ \textrm{or}~~
\begin{matrix}
(\slashed{p}- \slashed{\Sigma}_L^v) v_L = -M_\nu^\dagger v_R\\
(\slashed{p} - \slashed{\Sigma}_R^v) v_R = -M_\nu v_L\\
\end{matrix}
\label{EoM}
}
where $M_\nu$ is either Majorana or Dirac mass matrix and 
the self-energy correction $\slashed{\Sigma}$ takes the general form of
\dis{
\slashed{\Sigma} = \Sigma_\mu \gamma^\mu =
\left[ \Sigma_1 p_\mu  + \Sigma_2 k_\mu \right] \gamma^\mu 
\label{Sigmamu}  
}
both for $u_L$ and $u_R$, or $v_L$ and $v_R$.
Here $k_\mu$ is the four momentum of the background field and $\Sigma_{1,2}$ are functions of $p_\mu, k_\mu$ and the medium density.
The self-energy corrections of a particle and antiparticle are related by
\dis{
\Sigma^v_\mu(p)= -\Sigma^u_\mu(-p)=\Sigma_1^u(-p) p_\mu  - \Sigma_2^u(-p) k_\mu,
\label{u-v}
}
that is,  $\Sigma^v_{1}(p)=\Sigma^u_{1}(-p)$ and $\Sigma^v_2(p)= - \Sigma^u_2(-p)$, both for  
the left- and right-handed spinors. 
Thus the CP-violating effect appears through
\dis{
&\Sigma^u_\mu(p)  - \Sigma^v_\mu(p) \\
&\quad =[\Sigma_1^u(p) - \Sigma_1^u(-p) ]p_\mu +  [\Sigma_2^u(p) + \Sigma_2^u(-p) ]k_\mu
}
showing that the asymmetric contributions of $\Sigma_1$ and $\Sigma_2$ are odd and even under the change of $p\to -p$.

For the Majorana neutrino, we have the relation:
\dis{
u_R \equiv v_L^c \quad \textrm{and} \quad  v_R \equiv u_L^c,
}
and thus
\dis{
[\Sigma_R^u(p)]_\mu &= [\Sigma_L^v(p)  ]^T_\mu= -[ \Sigma_L^u (-p)]^T_\mu.
\label{SigmauR}
}

So far, all the equations are applicable to the multi-flavor case for which it is non-trivial to solve the equations of motion.
For a single flavor case with the mass $m_\nu$,  the solutions of the equations of motion for neutrinos can be found when 
the  determinant of the inverse propagator vanishes~\cite{Weldon:1982bn}:
\dis{
\textrm{Det}[ \slashed{L}  \slashed{R} - m_\nu^2] =& \textrm{Det}[ \slashed{R}  \slashed{L} - m_\nu^2 ] \\
= &L^2 R^2 -2 m_\nu^2 L\cdot R + m_\nu^4=0 \label{Det}
}
with $ L \equiv p-\Sigma_L$ and $ R \equiv p-\Sigma_R$,
which is applicable to both $u$ and $v$ spinors manifesting the standard relation: the negative energy solutions  for  the $u$ spinor correspond 
to the positive ones for the $v$ spinor. 
\\

In the rest frame of the medium, $k_\mu = m_{\phi} (1,\vec{0})$, Eq.~(\ref{Det}) is factorized to $(L\cdot R-m_\nu^2)^2=H^2$ allowing two types of solutions:
\dis{ \label{Det2}
 L\cdot R - m^2_\nu \pm H =0 .
 }
Here we defined $H\equiv L_{\bf p} R_0- R_{\bf p} L_0$ with  $L\equiv (L_0, \hat{\bf p} L_{\bf p})$ and $R\equiv (R_0, \hat{\bf p} R_{\bf p})$ introducing the notation of $p=(E,\vec{\bf p})$ and $\hat{\bf p} =\vec{\bf p}/|\vec{\bf p}|$.  More explicitly,  we have
\dis{ \label{Det3}
(E^2  &-  \bfp^2)(1-\Sigma_{1L})(1-\Sigma_{1R})-m^2_\nu
+m^2_\phi \Sigma_{2L}\Sigma_{2R} \\
&~-m_\phi (E\pm \bfp) \Sigma_{2L}(1-\Sigma_{1R})\\ 
&~-m_\phi (E\mp \bfp) \Sigma_{2R}(1-\Sigma_{1L})=0
}
where $\bfp\equiv |\vec{\bf p}|$.

While it is impossible to find an analytic solution to the dispersion equation in general, it is useful to
consider an approximate  dispersion relation and illustrate some basic properties in a model-independent way. More complete solutions will be drawn in the next section considering various limits of a specific model. 

When $|\Sigma_1|\ll 1$ and $m_\phi |\Sigma_2|\ll m_\nu$ or $\bfp$, it is straightforward to find the solution of \eq{Det2} up to the first order in perturbation:
\dis{
E^2_{\nu_1,\nu_2}  & \approx   \bfp^2 +m^2_\nu  +   m^2_\nu (\Sigma_{1L}^{(0)}+\Sigma_{1R}^{(0)} )\\
& +  {m_\phi} \left(   E_0 ( \Sigma_{2L}^{(0)}+\Sigma_{2R}^{(0)}) 
\pm  \bfp (\Sigma_{2L}^{(0)}-\Sigma_{2R}^{(0)} ) \right)
\label{Epert}
}
where we used the notation $\Sigma^{(0)} \equiv \Sigma(E_0,\vec{\bfp})$ with $E_0\equiv  \sqrt{{\bf p}^2 +m^2_\nu}$.  For a general fermion with four degrees of freedom, there appear four solutions of the dispersion relation.  Two positive-energy solutions  of \eq{Epert} describe the dispersion relations of two neutrino states  taking $\Sigma=\Sigma^u$, and two negative-energy solutions correspond to those of the anti-neutrino states  which are obtained with  $\Sigma=\Sigma^v$.
In the limit ${\bf p} \to \infty$ (or $m_\nu \to 0$), the last term of \eq{Epert} becomes
$2m_\phi \Sigma_{2L}$ or $2m_\phi \Sigma_{2R}$ representing the dispersion of $\nu_L$ or $\nu_R$.
In the opposite limit ${\bf p} \to 0$,  both $\nu_1$ and $\nu_2$ receive the same correction as expected
for the massive fermions at rest.\\

The dispersion relations in \eq{Det3}  or  (\ref{Epert})  can be adapted to the three types of neutrinos as follows.

i) {\bf Weyl neutrino} ($m_\nu=0$):\\
For a massless neutrino,  $\nu_L$ ($\nu_R$) is identified to $\nu$ ($\bar\nu$) and equivalently
\eq{Det3} is further decomposed into two seperate equations of motion 
\dis{
(E-\bfp)(1-\Sigma_{1L,R}) - m_\phi \Sigma_{2L,R} =0
}
corresponding to the positive energy solutions of $L^2=0$ and $R^2=0$
with $\Sigma_L=\Sigma_L^u$ and $\Sigma_R=\Sigma_R^u$, respectively. 
Thus,  we have the approximate solutions,
\begin{equation}
E_{\nu, \bar\nu} \approx {\bf p} + m_\phi \Sigma^{(0)}_{2L,2R},
\label{Epert_Weyl}
\end{equation}
which correspond to $E_{\nu_1, \nu_2}$ with $m_\nu=0$ in  \eq{Epert}. 
In an asymmetric medium, Weyl neutrinos will receive asymmetric correction corresponding to the difference $\Sigma_{2L}-\Sigma_{2R}$. \\

ii) {\bf  Majorana neutrino} ($\nu=\nu^c$):\\
For a Majorana neutrino in the relativistic limit, we can identify  $\nu_L$ ($\nu_R$) to  $\nu$ ($\bar\nu$), and thus it is useful to take $u_L$ and $u_R$ as independent spinors using the relation \eq{SigmauR}  \cite{Choi:2019zxy}. Then, applying \eq{Epert},  we get
\dis{
{E_{\nu, \bar\nu} } &\approx \sqrt{{\bf p}^2 +m^2_\nu}+{m^2_\nu  (\Sigma_{1L}^{(0)}+\Sigma_{1R}^{(0)} ) \over 2  \sqrt{{\bf p}^2 +m^2_\nu} } \\
&+ {m_\phi\over 2} \left( ({\Sigma_{2L}^{(0)}+\Sigma_{2R}^{(0)} ) \pm { {\bf p} (\Sigma_{2L}^{(0)}-\Sigma_{2R}^{(0)}) \over  \sqrt{{\bf p}^2 +m^2_\nu}} }\right) ,
\label{Epert_Maj}
}
where $\Sigma_{L,R}\equiv \Sigma^u_{L,R}$ used in \eq{Epert}.
Note that
$\Sigma_L \neq \Sigma_R$ in an asymmetric medium, and $\Sigma_L-\Sigma_R$ picks  the asymmetric contribution. 
In a symmetric medium ( $L=R$), the neutrino and anti-neutrino have the same dispersion relation:  $E_\nu = E_{\bar\nu}$ trivially from Eq.~(\ref{Det3}) or (\ref{Epert_Maj}).
\\

iii) {\bf Dirac neutrino}  ($\nu\neq \nu^c$ and $m_\nu \neq 0$):\\
The left- and right-handed components of a Dirac neutrino may interact differently with a medium leading to $\Sigma_L \neq \Sigma_R$.  In the simple case of $\Sigma_R^{u,v}=0$, the relation \eq{Det3} takes a much 
simpler form and \eq{Epert} leads to 
\dis{
{E_{\nu_1, \nu_2} }&\approx \sqrt{{\bf p}^2 +m^2_\nu}
+ {m^2_\nu  \Sigma_{1L}^{(0)} \over 2  \sqrt{{\bf p}^2 +m^2_\nu} } \\
&+ {m_\phi \over 2} \Sigma_{2L}^{(0)}  \left( 1 \pm { {\bf p}\over  \sqrt{{\bf p}^2 +m^2_\nu}}  \right) 
\label{Epert_Dir}
}
for the two neutrino states $\nu_{1,2}$ with $\Sigma_L=\Sigma_L^u$, 
 and the corresponding negative-energy solutions give us $E_{\bar\nu_1,\bar\nu_2}$ for the anti-neutrino states by replacing $\Sigma_L^u  \to \Sigma_L^v$.

\section{Neutrinos interacting with a medium of complex scalar}
\label{complex}
As a specific example, we consider a left-handed neutrino interacting with a background complex scalar $\phi$ and a fermion mediator $f$
through a Yukawa coupling $g$:
\dis{
 \label{Yukawa}
{\cal L}_{int} =g \bar f P_L \nu \phi + g^* \bar\nu P_R f \phi^* + m_{f}\bar{f} f ,
}
where $m_f$ is the Dirac mass.

\begin{figure}[!t]
\begin{center}
\begin{tabular}{c} 
 \includegraphics[width=0.4\textwidth]{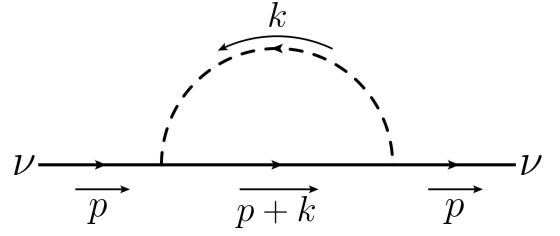}
     \end{tabular}
\end{center}
\caption{ Feynman diagrams for  the  forward scattering of  neutrino for the scenario of \eq{Yukawa}. }
\label{diagram}
\end{figure}

The self-energy correction of the left-handed neutrino is calculated  from the one-loop diagrams in Fig.~\ref{diagram}:
\dis{
\Sigma\!\!\!\!\slash\, = i|g|^2\int \frac{d^4 k}{(2\pi)^4}  \Delta_\phi(k)S_f(p+k),
}
where the bosonic and fermionic propagators in a finite density are given by 
\dis{
 \Delta_\phi(k) =& \frac{1}{k^2} - 2\pi i \delta(k^2-m_\phi^2) f_\phi(k),\\
S_f(q) =& (\slashed{q}+m_f) \left[ \frac{1}{q^2-m_f^2} + 2\pi i \delta(q^2-m_f^2) f_f(q)\right] .
}
Here $ f_\phi$ and $ f_f$ are distribution functions of scalar and fermion in the medium. 
In this work, we  assume that the medium is filled only by the complex scalar, that is, $f_f=0$
and 
\dis{
f_\phi (k) = [ \theta(k^0) n_\phi +  \theta(-k^0) n_{\phi^*}] (2\pi)^3 \delta^3(\vec{k}),
}
representing the number density of non-relativistic complex scalar in its rest frame:
\dis{
\int \frac{d^3 k }{(2\pi)^3} f_\phi(k) = \theta(k^0) n_\phi +  \theta(-k^0) n_{\phi^*} \nonumber
}
where $n_\phi$ and $n_{\phi^*}$ are the number densities of particle and anti-particle in the background, respectively.
The asymmetry of the complex scalar distribution is parameterized by $\epsilon$:
\dis{
n_\phi = \frac{N_\phi}{2} (1+\epsilon),\quad n_{\phi^*} = \frac{N_\phi}{2} (1-\epsilon).
}
where $N_\phi = n_\phi+n_{\phi^*}\equiv \rho_\phi/m_\phi$.

Using the relation
\dis{
\delta(k^2-m_\phi^2) = \frac{\delta(k^0-E_\phi) + \delta(k^0+E_\phi)}{2E_\phi} \nonumber
}
with $E_\phi \equiv \sqrt{|\vec{k}|^2+m_\phi^2} =m_\phi$ in the non-relativistic limit, one obtains the self-energy corrections:
\begin{eqnarray} \label{SigmaL}
\Sigma_{1L}^{u,v}(p)  & = & S(p) \pm \epsilon A(p),\\
\Sigma_{2L}^{u,v}(p)  & = & A(p) \pm \epsilon S(p), \nonumber
\end{eqnarray}
where $S(p)$ and $A(p)$ are the even  and odd functions of $p$ given by 
\dis{
S (p) \equiv \delta m_\nu^2
 \frac{p^2 +\mphi^2- m_f^2}{(p^2 +\mphi^2- m_f^2 )^2-4 \mphi^2 E^2} ,\\
A(p) \equiv  \delta m_\nu^2
 \frac{- 2m_\phi E}{(p^2 +\mphi^2- m_f^2 )^2-4 \mphi^2 E^2}.
 \label{SA}
}
Here we defined
\dis{
\delta m_\nu^2 \equiv |g|^2 \frac{N_\phi}{2m_\phi}.
\label{deltamnu2}
}
which is an effective mass-squared induced by the medium.
Note that the above result is consistent with the diagramatic calculation \cite{Choi:2019zxy}. 

 We are now ready to discuss explicit dispersion relations 
considering various limits of different phenomenological interests. 
In each case, we perform perturbative expansion and find 
order-by-oder solutions for the dispersion relation.

In the approximate relation (\ref{Epert}), the zeroth order expressions of $\Sigma^{(0)}$  can be written explicitly   in terms of $S_0$ and $A_0$ defined by
\dis{
S_0 \equiv \delta m_\nu^2
 \frac{m_\nu^2 +\mphi^2- m_f^2}{(m_\nu^2 +\mphi^2- m_f^2 )^2-4 \mphi^2 E_0^2} ,\\
A_0 \equiv  \delta m_\nu^2
 \frac{- 2m_\phi E_0}{(m_\nu^2 +\mphi^2- m_f^2 )^2-4 \mphi^2 E_0^2},
 \label{SA0}
}
with $E_0=\sqrt{\bfp^2+m_\nu^2}$. Then, the validity condition $|\Sigma_1^{(0)}|\ll1$ requires $|S_0|$, $| \epsilon A_0| \ll1$. 
Considering the medium of an ultra-light scalar field $\phi$, these quantities depend on the four energy scales: $\delta m^2_\nu, m^2_\nu, m_f^2$ and 
$2m_\phi \bfp$ ($m^2_\phi$ being the smallest). Then, the condition of $|S_0| \ll 1$, in particular, is realized in different physical situations
when either of $m_\nu^2, m_f^2, 2 m_\phi \bfp$ dominates the others. In each case, we will use Eq.~(\ref{Epert}) or (\ref{Epert_Maj}) to obtain the corresponding leading order correction. 

To obtain more precise dispersion relations, one needs to perform order-by-order expansion from the original equation \eq{Det} and  (\ref{Det2}) and find the solutions in each order. This is also necessary for the situation when $\delta m^2_\nu$ is the largest as it gives us $\Sigma_1 \simeq 1$.

 \subsection{Dispersion relation of  Weyl or Majorana neutrino}
 
 As described before, \eq{Det3} is applied to he Majorana neutrino with $\Sigma_{1L,R}=S\pm \epsilon A$ and $\Sigma_{2L,R}=A\pm \epsilon S$. Then,
the approximate dispersion relation (\ref{Epert_Maj}) takes the form:
  \dis{
  E_{\nu,\bar{\nu}} & \simeq\sqrt{\bfp^2+m_\nu^2} + \frac{m_\nu^2 S_0}{\sqrt{\bfp^2+m_\nu^2} }\\
   & +\mphi \left( A_0 \pm  \epsilon \frac{ \bfp   S_0 }{\sqrt{\bfp^2+m_\nu^2} }\right),
  \label{Epert_WM}
  }
 where the sign $+$ ($-$) is for neutrino (antineutirno).
  In the CP asymmetric background ($\epsilon\neq0$), the neutrino and anti-neutrino energies are split by the asymmetric term $\epsilon S_0$
  which vanishes in the limit $\bfp\to 0$ for the Majorana neutrino ($m_\nu \neq 0$). 
  The above relation can be applicable to the Weyl neutrino taking the limit $m_\nu=0$ reproducing (\ref{Epert_Weyl}). 
  Let us now discuss four different situations introduced above.

 \subsubsection{Decoupling limit:  $m_f^2$ is the largest}
In the limit of $m_f^2 \to \infty$, the leading contribution is obtained from \eq{Epert_Maj} by putting $S_0\approx -\delta m_\nu^2/m_f^2 $ and ignoring  the next order term $A_0 \sim {\cal O}(1/m_f^4)$:
 \dis{
 E_{\nu,\bar{\nu}} \simeq& \sqrt{\bfp^2+\mnu^2} - \frac{\dmnu}{\mf}\frac{(m_\nu^2\pm\epsilon \mphi \bfp)}{\sqrt{\bfp^2+\mnu^2} }.
 }
 This shows that the rest mass is slightly modified to $m_\nu (1-\delta m^2_\nu/m_f^2)$ and the dispersion relation in 
the  high-momentum limit $ m_\nu \ll \bfp \ll \mf/(2\mphi)$ is given by
\dis{
E_{\nu,\bar{\nu}} \simeq \bfp + \frac{m_\nu^2}{2\bfp} \left( 1- 2\frac{\dmnu}{\mf} \right)\mp\frac{\dmnu}{\mf}  \epsilon \mphi,
}
which contains a tiny mass split between the neutrino and anti-neutrino.
Notice that the second term describing the neutrino oscillation is corrected by a small factor $(1-2\delta m^2_\nu/m_f^2)$
consistently with the rest mass correction.

Taking $m_\nu=0$, one obtains the dispersion relation of the Weyl neutrino which contains only the asymmetric mass term at this order.

 \subsubsection{Heavy neutrino limit:  $m_\nu^2$ is the largest}
 In this case, we take $S_0 \approx \frac{\delta m_\nu^2}{m^2_\nu}\left(1+\frac{m_f^2}{m_\nu^2}\right) \gg A_0$ including the smaller corrections with $m_f^2$ to get
  \dis{
 E_{\nu,\bar{\nu}} & \simeq  \sqrt{\bfp^2+m_\nu^2} + \frac{\dmnu}{ \sqrt{\bfp^2+m_\nu^2} }\\ 
 & + \frac{\dmnu}{m^2_\nu}\frac{(m_f^2\pm\epsilon \mphi \bfp)}{\sqrt{\bfp^2+\mnu^2} },
 }
 which gives
 \dis{ \label{Mheavy}
E_{\nu,\bar{\nu}} \simeq \bfp + \frac{m_\nu^2 +2 \delta m_\nu^2}{2\bfp}
 \pm \frac{\dmnu}{m_\nu^2}  \epsilon \mphi
}
in the relativistic limit $\bfp\gg m_\nu$. Here the rest mass is changed to $m_\nu (1+\delta m^2_\nu/m_\nu^2)$ which gives   
the neutrino oscillation term  shifted by a small correction  $2\delta m^2_\nu$.

\subsubsection{High momentum limit: $2 m_\phi \bfp$ is the largest}
Although we consider typically an ultra-light scalar $\phi$ in the medium,  $2m_\phi \bfp$ can be much larger than the other mass scales
$m_f^2, \delta m_\nu^2$ or $m^2_\nu$ realized in various neutrino experiments with $\bfp\gtrsim$ MeV. In this case, we have 
$A_0 \approx  \delta m^2_\nu/ 2 m_\phi \bfp \gg S_0$ in the leading order. In order to get the next leading order terms consistently,
we expand the original equation \eq{Det3} in (inverse) powers of $2m_\phi \bfp$, and obtain the solution of $E^2-\bfp^2$ up to ${\cal O}(1/2m_\phi \bfp)$. 
Taking then the relativistic limit ($\bfp^2 \gg 2 m_\phi \bfp\gg m_\nu^2$) one finds
  \dis{
  E_{\nu,\bar{\nu}}  \simeq \bfp+ \frac{m_\nu^2+ \dmnu}{2\bfp}   \mp \epsilon \frac{\dmnu (m_\nu^2-\mf)}{4 \mphi \bfp^2} ,
   \label{highpE}
 }
 which is in fact obtained from \eq{Epert_WM} keeping the $\epsilon S_0$ term.
 This shows that the neutrino oscillation can occur through the medium-induced mass-squared $\delta m^2_\nu$ even for the Weyl neutrino ($m_\nu = 0$) , and the CP violating effect may appear in neutrino oscillations through the correction: $\mp \epsilon \delta m_\nu^2(m^2_\nu-m_f^2)/2m_\phi \bfp$~\cite{Choi:2019zxy}.
 
 \subsubsection{High density limit: $\delta m^2_\nu$ is the largest}
 This is the case where the perturbative solution \eq{Epert} is not valid. The key feature is that $\delta m^2_\nu \gg 2 m_\phi \bfp$, $m_f^2 \gg m_\phi^2$ although  $\bfp^2$ (or $m^2_\nu$) can be larger than $\delta m^2_\nu$.
Expanding \eq{Det2} in powers of the small quantities, $2m_\phi \bfp$ and $m_f^2$ ($\gg m^2_\phi$), 
one obtains a peculiar dispersion relation up to the first order of perturbation:
\dis{ \label{Mrest}
E_{\nu, \bar\nu}^2 \approx \bfp^2 +m_M^2 + 2 \delta m^2_\nu \frac{m_f^2 \pm \epsilon m_\phi \bfp}{m_M^2+\delta m^2_\nu} ,
}
where $m_M^2 \equiv  \frac12\left( m_\nu^2 + 2\dmnu +\sqrt{m_\nu^4 +4m_\nu^2 \dmnu} \right)$, or
$m_M=(m_\nu + \sqrt{m_\nu^2+4 \delta m^2_\nu})/2$.
One finds that $m_M^2 \approx m^2_\nu + 2 \delta m^2_\nu$ for $m^2_\nu \gg \delta m^2_\nu$
and thus \eq{Mrest} reproduces \eq{Mheavy} in the limit of  $m_\nu^2 \gg \delta m^2_\nu$.

It is remarkable that there appears a non-vanishing  contribution  $\delta m_\nu^2$ even in the Weyl limit $m_\nu \to 0$. However, we have $\Sigma_1 \approx 1$ in this case, and thus the one-loop result appears not reliable. 
A non-perturbative calculation would be required to obtain the  precise dispersion relation in the high density medium.

 \subsection{Dispersion relation of Dirac neutrino}
 When  the Yukawa interaction \eq{Yukawa} is applied to Dirac neutrinos, 
only the propagator of the left-handed neutrino gets modified and 
the corresponding self-energy corrections for the $u$- and $v$-spinors are 
given by Eq.~(\ref{SigmaL}) while $\Sigma_R\equiv 0$.
Thus, \eq{Epert_Dir} becomes 
\dis{
{E_{\nu_1, \nu_2} }&\approx \sqrt{{\bf p}^2 +m^2_\nu}
+ {m^2_\nu (S_0+\epsilon A_0)  \over 2  \sqrt{{\bf p}^2 +m^2_\nu} } \\
&+ {m_\phi \over 2} (A_0+\epsilon S_0)   \left( 1 \pm { {\bf p}\over  \sqrt{{\bf p}^2 +m^2_\nu}}  \right) ,
\label{Epert_Dirac}
}
 and $E_{\bar\nu_1,\bar\nu_2}$ is obtained by  
$\epsilon\to -\epsilon$.
Note again that $E_{\nu_2} \to \bfp$ for $m_\nu\to 0$, that is, $\nu_2$ corresponds to $\nu_R$. 

Repeating the previous discussions, we will consider the cases that each of the  mass scales, 
 $m_f^2$, $m_\nu^2$, $2m_\phi \bfp$ or $\delta m_\nu^2$,  dominates.

 \subsubsection{Decoupling limit: $m^2_f$ is the largest}
Putting again  $S_0 \approx -\delta m^2_\nu/m_f^2 \gg A_0$ in \eq{Epert_Dirac}, we get
 \dis{ \label{DD1}
   E_{\nu_1,\nu_2} & \simeq \sqrt{\bfp^2+m_\nu^2} \\
  & - \frac{\dmnu}{2\mf}  \left(
   \frac{\mnu^2 + \epsilon \mphi \left[ \sqrt{\bfp^2+m_\nu^2} \pm  \bfp \right]  }{\sqrt{\bfp^2+m_\nu^2} } \right).
}
Thus, the rest mass is corrected to $
\mnu \left( 1-  \frac{\dmnu}{2\mf} \right) - \epsilon \mphi \frac{\dmnu}{2\mf}$
for both neutrino states $\nu_{1,2}$.
In the relativistic limit, however, we have
\dis{ \label{DD3}
  E_{\nu_1,\bar{\nu}_1} & \approx  \bfp + \frac{m^2_\nu}{2\bfp} \left( 1- \frac{\dmnu}{\mf} \right)  \mp \frac{\dmnu}{\mf}  \epsilon \mphi,\\
 E_{\nu_2,\bar{\nu}_2} & \approx  \bfp + \frac{m^2_\nu}{2\bfp} \left( 1- \frac{\dmnu}{\mf} \right),
}
showing a difference between the $\nu_{1,2}$ states.

\subsubsection{Heavy neutrino limit:  $m_\nu^2$ is the largest}
In this case, we use $S_0\approx \frac{\delta m_\nu^2 }{m_\nu^2 }\left(1+\frac{m_f^2}{m_\nu^2}\right) $ and $A_0\approx - \delta m^2 ( 2m_\phi  \sqrt{\bfp^2+m_\nu^2})/m_\nu^4$ keeping also the next leading order corrections which will be justified later.
Then, we obtain the dispersion relation
 \dis{\label{HeavyDirac}
   E_{\nu_1,\nu_2} &  \approx \sqrt{\bfp^2+m_\nu^2} + \frac{\delta m^2_\nu}{2\sqrt{\bfp^2+m_\nu^2} } \\
  & + \frac{\dmnu}{2m_\nu^2}  \left(
   \frac{m_f^2-\epsilon \mphi \left[ \sqrt{\bfp^2+m_\nu^2} \mp  \bfp \right]  }{\sqrt{\bfp^2+m_\nu^2} } \right),
}
which gives the rest mass correction: $m_\nu (1+\delta m^2_\nu/ 2 m_\nu^2) - \epsilon \mphi \frac{\dmnu}{2m_\nu^2}$. In the relativisitic limit, an opposite  behavor appears compared to the decoupling limit:
\dis{ \label{HMD2}
  E_{\nu_1,\bar{\nu}_1} &\approx  \bfp + \frac{m_\nu^2 +\dmnu}{2\bfp}, \\
 E_{\nu_2,\bar{\nu}_2} & \approx  \bfp + \frac{m_\nu^2+\dmnu}{2\bfp} 
 \mp \frac{\dmnu}{m^2_\nu}  \epsilon \mphi,
}
showing no asymmetric correction to the $\nu_1 \approx \nu_L$ states. 
Note the difference of factor 2  compared with the previous Majorana case.

\subsubsection{High momentum limit: $2 m_\phi \bfp$ is the largest}
For this case, we  again expand \eq{Det3} in powers of $\{ \delta m_\nu^2, m_\nu^2, m_f^2\}/2m_\phi \bfp \gg m_\phi^2/2m_\phi\bfp$ to obtain
\dis{
  E_{\nu_1,\bar{\nu}_1} & \simeq  \bfp+     \frac{m_\nu^2+ \delta m^2_\nu}{2\bfp} 
   \pm \epsilon  \frac{\delta m^2_\nu}{2\bfp}   \frac{m_f^2}{2\mphi \bfp},  \\
  E_{\nu_2,\bar{\nu}_2} &\simeq   \bfp+   \frac{m_\nu^2}{2\bfp}\left( 1\pm \epsilon \frac{\dmnu}{2\mphi \bfp} \right),
  \label{HighEDirac}
}
taking the relativistic limit. 
Notice the neutrino oscillation term  $(m_\nu^2+\delta m_\nu^2)/2\bfp$  for the left-handed neutrino,
which is the same as the Majorana case, compared to $m_\nu^2/2\bfp$  for the right-handed neutrino.
There is also a difference in the asymmetric terms.
 
 \subsubsection{High density limit: $\delta m^2_\nu$ is the largest}
Taking a power series of  \eq{Det3} in terms of $2m_\phi E_0$, $2m_\phi \bfp$, and $m_f^2$ ($\gg m_\phi^2$) which are assumed to be much smaller than $\delta m^2_\nu$ (and $m_\nu^2$),  we find the dispersion relation up to the first order in perturbation:
\dis{
E_{\nu_1, \nu_2}^2 \approx \bfp^2 + m_D^2  +\dmnu  \frac{m_f^2 - \epsilon \mphi (E_0 \mp  \bfp)}{ m_D^2 },
}

where $m_D^2 \equiv m^2_\nu+\dmnu$ and $E_0 \equiv \sqrt{ \bfp^2 +m_D^2}$. Taking the limit $m^2_\nu \gg \delta m^2_\nu$, one reproduces \eq{HeavyDirac} confirming the validity of the approximation.
One can see that the difference between the left- and right-handed neutrinos appears in the asymmetric term which vanishes for $\bfp\to 0$.  It is worthwile to note again the difference in the rest mass ocrrections, $m_D^2$ vs.\ $m_M^2$, between the Dirac and Majorana neutrinos, which will lead to a drastically different oscillation terms, $m_D^2/2\bfp$ vs.\ $m_M^2/2\bfp$, in the relativistic limit.

\section{Implications}
\label{implications}
One of the important consequences of the medium effect is the appearance of the neutrino oscillation term
\begin{equation}
 E\sim {m_\nu^2 + \delta m_\nu^2 \over 2 \bfp},
 \end{equation}
in the ultra high-momentum limit.  This implies that the observed neutrino oscillations can be due to the flavor mixing by the rest mass (Majorana or Dirac) and/or by the medium-induced ``mass-squared'' which does not involve chirality-flip or lepton-number violation. As a consequence, 
(massless) Weyl neutrinos can also oscillate in a medium~\cite{Choi:2019zxy}.
 
A natural candidate of such a medium is the ultra-light dark matter (DM) background which produces the effective neutrino mass-squared: $\delta m^2_\nu \propto \rho_{DM}/m^2_{DM}$. In this case, the DM density increases in the early universe and thus neutrinos undergo a high-density regime discussed in the previous section.  At a redshift $z$, we have $\rho_{DM}(z) = \rho^0_{DM} (1+z)^3$ where $\rho^0_{DM}$ is the average DM density at present,  and thus $\delta m^2_\nu$ becomes huge in the early Universe. 
Compared to the local DM density producing the observed neutrino mass-squared, $\delta m^2_{local}$, 
one has
\begin{equation} 
\delta m^2_\nu (z)= \delta m^2_{local} \frac{\rho^0_{DM}}{\rho^{local}_{DM}}(1+z)^3 \approx 6650
 \left(z\over {1100}\right)^3 \,  \delta m^2_{local},
\end{equation}
where we used $\rho^{local}_{DM} =0.3\gev/\cm^{3}$ and $\rho^0_{DM}=1.5\times 10^{-6}\gev/\cm^{3}$.
The energy density provided by the neutrinos during the decoupling time is constrained by the CMB measurement putting the upper bound on the neutrino mass $\sim 0.1$ eV~\cite{Aghanim:2018eyx}.  This implies
\begin{equation}
 \delta m^2_{local  } \lesssim 10^{-6} {\rm eV}^2,
\end{equation}    
which is too small to explain both the atmospheric and solar neutrino oscillations. Thus,
the current density of DM interacting with neutrinos is required to be produced well after the decoupling time. To meet this unconventional demand, one could think of an early DM density consisting of another particles decaying very late to produce the neutrino-favorite DM background.  Otherwise, it puts a strong limit on the effective neutrino mass induced by the current DM distribution.  

\begin{figure}[!t]
	\begin{center}
		\begin{tabular}{c} 
			\includegraphics[width=0.44\textwidth]{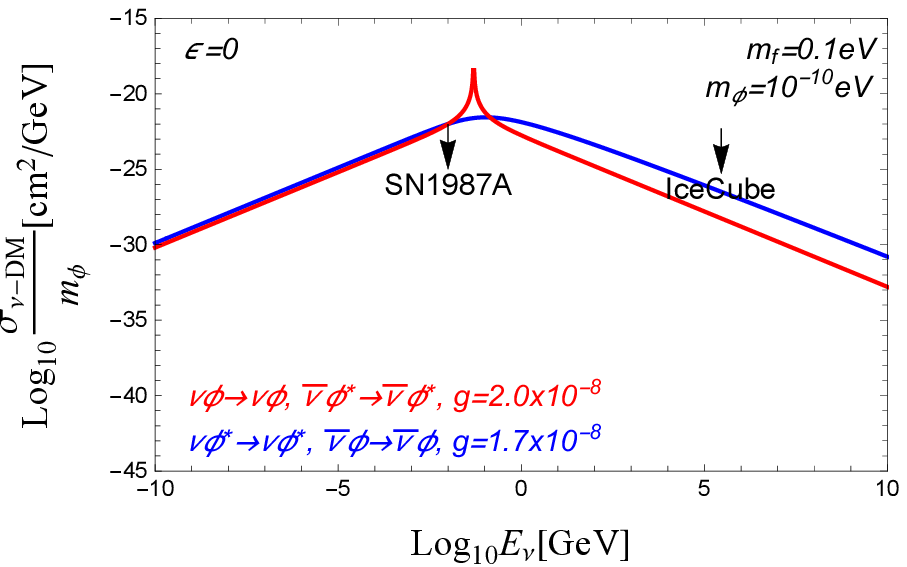}
			\\
		~	\includegraphics[width=0.42\textwidth]{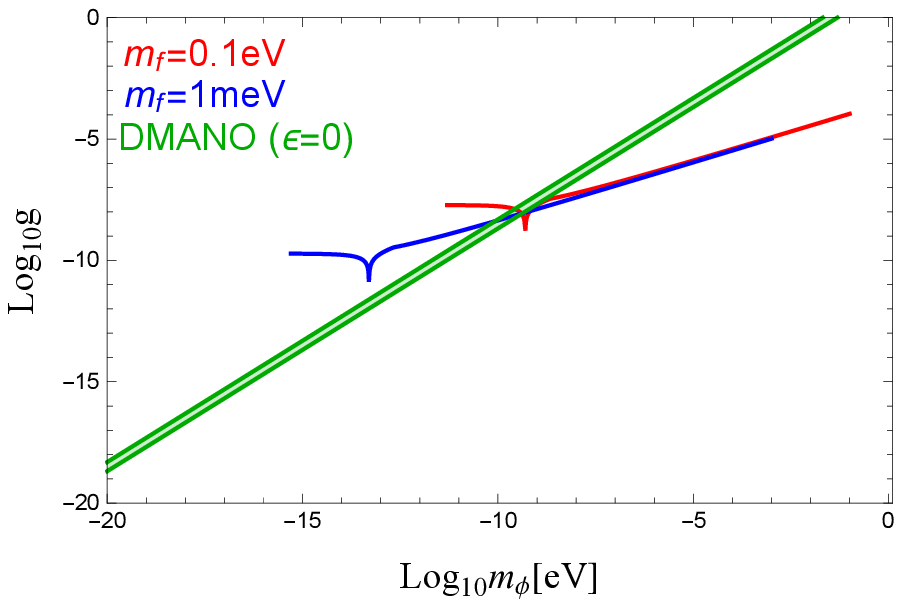}
		\end{tabular}
	\end{center}
	\caption{(Top) The elastic scattering cross-sections of neutrinos with DM vs.\ the neutrino energy in comparison with  the SN 1987A  and IceCube-170922A  data.  (Bottom) The allowed region for the DM-assisted neutrino oscillation  (green band) constrained by elastic scattering with DM (the lower region of the line is allowed, blue for $m_f=10^{-3}\ev$) and red for $m_f=0.1\ev$.} 
	\label{Bounds}
\end{figure}

At present, the scattering cross-section between neutrinos and DM  is constrained by astrophysical neutrino observations from SN1987A and IceCube-170922A  for the neutrino energy around $10 \mev$ and $300\tev$, respectively~\cite{Choi:2019ixb}:
\dis{
\sigma_{\nu-DM} / m_{\phi} \lesssim 10^{-22} \cm^2/\gev.
}
In our scenario with the  Yukawa interaction \eq{Yukawa}, 
the scattering cross section between neutrino and dark matter at high energy is
\dis{
\sigma_{\nu-\phi} \simeq  \frac{g^4}{32\pi \mphi  E_\nu },
}
and thus  it requires  
\dis{
\mphi > 2 \times 10^6 g^2 \, \ev.
}
On the other hand,  the effective neutrino mass-squared induced by the local DM distribution can describe the observed neutrino oscillations for
\dis{
\mphi \approx 0.03\, g \ev  \bfrac{2.5\times 10^{-3}\ev^2}{\delta m_\nu^2}^{1/2},
}
obtained from $\delta m^2_\nu = g^2 \rho^{local}_{DM} /2m_\phi$ (\ref{deltamnu2}).
Combing these two conditions,  we get 
\dis{
g& \lesssim 1.5\times 10^{-8}  \bfrac{2.5\times 10^{-3}\ev^2}{\delta m_\nu^2}^{1/2}, \\
 \mphi & \lesssim 4.5\times 10^{-10}\ev  \bfrac{2.5\times 10^{-3}\ev^2}{\delta m_\nu^2}^{1/2} .
}
The mediator mass is bounded from below  $m_\phi < m_f$ for the stability of $\phi$, and the upper bound
\dis{
m_f \ll 1 \ev  \bfrac{2.5\times 10^{-3}\ev^2}{\delta m_\nu^2}^{1/4} \bfrac{E_\nu}{\rm GeV}^{1/2}\\
}
is required for $2 m_{\phi} E_\nu \gg m_f^2$.

In  Fig.~\ref{Bounds}, we present the result of more precise calculation. The top panel shows  the astrophysical bounds on $\sigma_{\nu-DM}/m_{\phi}$ as a function of the neutrino energy for $m_\phi=10^{-10}$eV, $m_f=0.1$eV, and $\epsilon=0$. The upper bounds on the Yukawa coupling are shown in blue and red depending on the types of neutrino-DM interactions.
In the bottom panel, the green band shows the region of  the DM mass and Yukawa coupling required by the atmospheric neutrino oscillation for the effective neutrino mass induced by the local DM distribution in order to solely resolve the current observations of neutrinos. 
The blue ($m_f=10^{-3}\ev$) and red ($m_f=0.1\ev$) solid lines represent the biggest coupling constant limited by the conditions:  $m_\phi < m_f$ and  
$m_\phi > m^2_f/2E_\nu $ with $E_\nu = 1$ GeV. The lower region of the line is allowed.

\section{Conclusion}
\label{conclusions}
Dispersion relations of Weyl, Majorana and Dirac neutrinos propagating in a complex scalar medium are analyzed to study the medium effect in neutrino oscillations. Considering various limits of mass and energy scales in the model, we performed perturbative calculation to obtain the dispersion relation specific to each case. As a result, one can see how the medium effect modifies the neutrino oscillation term and  contributes to the CP asymmetric correction. It is remarkable that the corrections to the mass-squared are different for Majorana and Dirac neutrinos in the high-density limit, and thus leading to different corrections to the neutrino oscillation terms. 

Considering the medium of ultra-light scalar DM, one finds that neutrinos become  ``too heavy'' in the early universe and thus the cosmological scenario for neutrino-DM interaction is highly restricted. 
On the other hand, the astrophysical neutrino observations put additional constraints on the model parameters in the high-momentum limit.   

\begin{acknowledgments}
K.-Y.C. was supported by the National Research Foundation of Korea (NRF) grant funded by the Korea government (MEST) (NRF-2019R1A2B5B01070181). J. Kim was supported by a KIAS Individual Grant (PG074201) at Korea Institute for Advanced Study. This project has received support from the European Union’s Horizon 2020 research and innovation programme under the Marie Skłodowska-Curie grant agreement No 690575.
\end{acknowledgments}


\end{document}